\documentclass[conference]{template/IEEEtran}
\IEEEoverridecommandlockouts

\usepackage[T1]{fontenc}
\usepackage[noadjust,sort]{cite}
\usepackage{amsmath,amssymb,amsfonts}
\usepackage{algorithmic}
\usepackage{graphicx}
\usepackage{textcomp}
\usepackage{xcolor}
\def\BibTeX{{\rm B\kern-.05em{\sc i\kern-.025em b}\kern-.08em
    T\kern-.1667em\lower.7ex\hbox{E}\kern-.125emX}}

\usepackage[american]{babel}
\usepackage[per-mode=symbol,mode=text,detect-all]{siunitx}
\DeclareSIUnit[quantity-product = {}]{\percent}{\char`\%}
\usepackage{microtype}
\usepackage{booktabs}
\usepackage{multirow}
\usepackage{calc}
\usepackage[inline]{enumitem}
\usepackage{acro}
\usepackage{fancyvrb}
\usepackage[hidelinks]{hyperref}
\usepackage[capitalize,nameinlink,noabbrev]{cleveref}
\usepackage{enumitem}
\usepackage{xspace}
\usepackage{balance}
\usepackage{flushend}

\DeclareAcronym{bmftr}{short=BMFTR, long={Federal Ministry of Research, Technology and Space}}
\DeclareAcronym{ae}{short=AE, long=Artifact Evaluation}
\DeclareAcronym{ai}{short=AI, long=Artificial Intelligence}
\DeclareAcronym{api}{short=API, long=Application Programming Interface}
\DeclareAcronym{gpu}{short=GPU, long=Graphics Processing Unit}
\DeclareAcronym{cpu}{short=CPU, long=Central Processing Unit}
\DeclareAcronym{hpc}{short=HPC, long=High Performance Computing}
\DeclareAcronym{ids}{short=IDS, long=Intrusion Detection System}
\DeclareAcronym{llm}{short=LLM, long=Large Language Model}
\DeclareAcronym{pca}{short=PCA, long=Principal Component Analysis}
\DeclareAcronym{ui}{short=UI, long=user interface}
\DeclareAcronym{gui}{short=GUI, long=graphical user interface}
\DeclareAcronym{iot}{short=IoT, long=Internet of Things}
\DeclareAcronym{cps}{short=CPS, long=Cyber-Physical System}
\DeclareAcronym{ics}{short=ICS, long=Industrial Control System}

\newcommand{\ie}{i.e.\@\xspace}

\newcommand{\eg}{e.g.\@\xspace}

\newcommand{\cf}{cf.\@\xspace}
\newcommand{\etal}{et~al.\@\xspace}

\let\tmpcite\cite
\renewcommand{\cite}{%
  \ifhmode%
    \ifdim%
      \lastskip>0pt%
      \unskip~%
    \else%
    \fi%
  \fi%
  \tmpcite%
}

\newcommand{\rate}{\textsc{Rate}\xspace}
\newcommand{\prep}{\textsc{Prepare}\xspace}
\newcommand{\ass}{\textsc{Assess}\xspace}

\newcommand{\comp}{\textsc{Complement}\xspace}

\newcommand{\olsdata}{\textsc{Olszewski-Study}\xspace}
\newcommand{\arpdata}{\textsc{Arp-Study}\xspace}

\newcommand{\readme}{Readme\xspace}

\newcommand{\llamathreetwothreebinstruct}{\textsc{Llama-3.2-3B-Instruct}\xspace}
\newcommand{\llamathreetwothreebinstructfootnote}{\url{https://www.llama.com/docs/model-cards-and-prompt-formats/llama3_2/}}
\newcommand{\gptfouromini}{\textsc{gpt-4o-mini}\xspace}
\newcommand{\gptfourominifootnote}{\url{https://platform.openai.com/docs/models/o4-mini}}

\newcommand{\issue}[1]{\emph{\textbf{Issue:} #1}}

\newlist{trianglelist}{itemize}{1}
\setlist[trianglelist,1]{nosep,label=$\blacktriangleright$,align=right,itemindent=\parindent+\widthof{~}+\widthof{$\blacktriangleright$},labelsep=\widthof{~},labelwidth=\widthof{$\blacktriangleright$},leftmargin=0pt}

\usepackage{tikz}
\newcommand\copyrighttext{
  \footnotesize
  \textcopyright 2025 IEEE.
  Personal use of this material is permitted.
  Permission from IEEE must be obtained for all other uses, in any current or future media, including reprinting/republishing this material for advertising or promotional purposes, creating new collective works, for resale or redistribution to servers or lists, or reuse of any copyrighted component of this work in other works.
}
\newcommand\copyrightnotice{%
  \begin{tikzpicture}[remember picture,overlay]
    \node[anchor=south,yshift=25pt] at (current page.south) {\parbox{\dimexpr\textwidth-\fboxsep-\fboxrule\relax}{\copyrighttext}};
  \end{tikzpicture}
}

\begin{document}

\bstctlcite{IEEEexample:BSTcontrol}

\title{Supporting Artifact Evaluation with LLMs:\\A Study with Published Security Research Papers}


\author{%
  \IEEEauthorblockN{%
    David Heye\IEEEauthorrefmark{1}\textsuperscript{,}\IEEEauthorrefmark{2},
    Karl Kindermann\IEEEauthorrefmark{1}\textsuperscript{,}\IEEEauthorrefmark{2},
    Robin Decker\IEEEauthorrefmark{1}\textsuperscript{,}\IEEEauthorrefmark{2},
    Johannes Lohmöller\IEEEauthorrefmark{1},\\
    Anastasiia Belova\IEEEauthorrefmark{3},
    Sandra Geisler\IEEEauthorrefmark{3},
    Klaus Wehrle\IEEEauthorrefmark{1},
    Jan Pennekamp\IEEEauthorrefmark{1}%
  }%
  \IEEEauthorblockA{%
    \IEEEauthorrefmark{1}\textit{Communication and Distributed Systems}, RWTH Aachen University, Germany $\cdot$
    \{lastname\}@comsys.rwth-aachen.de\\
    \IEEEauthorrefmark{3}\textit{Data Stream Management and Analysis}, RWTH Aachen University, Germany $\cdot$
    \{lastname\}@dbis.rwth-aachen.de%
  }%
  \thanks{\IEEEauthorrefmark{2} These authors contributed equally to this work.}
}

\maketitle

\makeatletter
\renewcommand{\subsubsection}{\@startsection{subsubsection}{3}{\parindent}{0ex plus 0.1ex minus 0.1ex}{0ex}{\normalfont\normalsize\itshape\bfseries}}
\makeatother

\begin{abstract}
\Ac{ae} is essential for ensuring the transparency and reliability of research, closing the gap between exploratory work and real-world deployment is particularly important in cybersecurity, particularly in \acs{iot} and \acsp{cps}, where large-scale, heterogeneous, and privacy-sensitive data meet safety-critical actuation.
Yet, manual reproducibility checks are time-consuming and do not scale with growing submission volumes.
In this work, we demonstrate that \acp{llm} can provide powerful support for \acs{ae} tasks:
\begin{enumerate*}[label=(\roman*)]
\item text-based reproducibility rating,
\item autonomous sandboxed execution environment preparation, and
\item assessment of methodological pitfalls.
\end{enumerate*}
Our reproducibility-assessment toolkit yields an accuracy of over \qty{72}{\percent} and autonomously sets up execution environments for \qty{28}{\percent} of runnable cybersecurity artifacts.
Our automated pitfall assessment detects seven prevalent pitfalls with high accuracy ($\mathbf{F_1} > \text{\qty{92}{\percent}}$).
Hence, the toolkit significantly reduces reviewer effort and, when integrated into established \acs{ae} processes, could incentivize authors to submit higher-quality and more reproducible artifacts.
\Acs{iot}, \acs{cps}, and cybersecurity conferences and workshops may integrate the toolkit into their peer-review processes to support reviewers' decisions on awarding artifact badges, improving the overall sustainability of the process.
\end{abstract}

\acresetall{}


\section{Introduction}
\label{sec:introduction}

\copyrightnotice{}
The rapid evolution of cyber threats poses a significant challenge for maintaining resilience across many networked domains, including \ac{iot} and \ac{cps} deployments, industrial control systems, connected vehicles, and smart cities.
Recent reports by the World Economic Forum \cite{World-Economic-Forum2025Global} and the European Union Agency for Cybersecurity \cite{European-Union-Agency-for-Cybersecurity2024ENISA} demonstrate that adversaries not only refine established attack vectors but also exploit emerging technologies, particularly \ac{ai}, to evade traditional defenses.
In response, the volume and complexity of security research have grown rapidly \cite{Olszewskietal2023Get}.
Yet a critical gap persists between proof-of-concept prototypes for research evaluation and solutions that are mature and robust enough for real-world deployment \cite{Olszewskietal2023Get}.
This gap undermines confidence in published results and obstructs the translation of academic advances into practical solutions.

To foster trust and accelerate the technological transfer, the academic community increasingly adopts reproducibility badges and performs \ac{ae} within the peer-review process \cite{Athanassoulisetal2022Artifacts,Olszewskietal2023Get,Krishnamurthi2014About}.
These processes require authors to submit code, data, and instructions, which independent reviewers use to verify claimed results.
However, \ac{ae} is labor-intensive, depends on volunteers with specialized expertise, and struggles to keep pace with the rising submission rate in cybersecurity conferences and workshops \cite{Olszewskietal2023Get}.

\Acp{llm} have demonstrated remarkable natural language understanding, code synthesis, and knowledge extraction capabilities.
In cybersecurity contexts, \acp{llm} have been applied to intrusion and anomaly detection \cite{Zhangetal2024Large}, secure coding assistance \cite{Sandovaletal2023Lost}, and automated penetration testing \cite{Happeetal2023Getting}.
Simultaneously, some researchers are looking into improving conventional peer review with \acp{llm} \cite{Caoetal2025CSPaper,Kuznetsovetal2024What,Zhuetal2025DeepReview:,Yeetal2024Are,Idahletal2025OpenReviewer:,Sunetal2024MetaWriter:,Huangetal2025PaperEval:}.
In this paper, we explore a new dimension of their utility: supporting and automating \ac{ae} for cybersecurity research.
In light of the growing number of submissions at cybersecurity venues, we aim to provide automated support for reviewers of scientific contributions to improve the scalability of \ac{ae}.
We introduce an \ac{llm}-driven toolkit that analyzes paper texts and accompanying artifacts to
\begin{enumerate*}[label=(\roman*)]
\item extract reproducibility indicators,
\item detect potential inconsistencies between claims and submitted artifacts, and
\item identify common pitfalls in experimental design and evaluation.
\end{enumerate*}
By embedding these capabilities into the peer-review workflow, we aim to improve both the scalability and consistency of the \ac{ae} process.

\textbf{Contributions.}
We propose a three-step \ac{llm}-driven toolkit to partially automate reproducibility assessments:
\begin{trianglelist}
\item \rate{}:
  Our \ac{llm}-based method that conceptualizes reproducibility via concept vectors extracted from the model's hidden states achieves a recall of almost \qty{95}{\percent}, allowing the automatic discarding of non-reproducible submissions.
\item \prep{}:
  Our \ac{llm}-agent framework automatically sets up and runs submitted artifacts in sandboxed environments, preparing nearly \qty{30}{\percent} of manually reproducible submissions and offering supporting hints for all others.
\item \ass{}:
  By repurposing the concept of the \rate{} stage, we reliably identify many design and evaluation pitfalls in security contributions, with an accuracy of \qty{>90}{\percent}.
\item \emph{Integrated pipeline}:
  A combination of these stages into a unified \ac{ae} workflow, which balances computational cost with assessment accuracy, correctly classifies more than \qty{72}{\percent} of the papers in our dataset regarding their reproducibility.
\end{trianglelist}

\textbf{Open Science.}
We have published our code on GitHub \cite{Comsys2025ArtifactEvaluationLLMSupport}.

\textbf{Organization.}
The remainder of this paper is structured as follows.
\Cref{sec:background} provides foundations and introduces recent works on reproducibility and relevant \ac{ai} techniques, particularly focusing on cybersecurity.
\Cref{sec:repr-pipel} details our three-stage \ac{llm}-driven pipeline.
\Cref{sec:evaluation} describes our implementation and empirical evaluation on a curated dataset of hundreds of security research papers.
\Cref{sec:discussion} discusses findings, lessons learned, and directions for future research before we conclude in \cref{sec:conclusion}.

\section{Background and Related Work}
\label{sec:background}

\Ac{ae} at security conferences has become vital for ensuring the transparency and reliability of research, fostering a collaborative environment among researchers and experts.
Cybersecurity presents unique challenges for \ac{ae}, often involving rapidly evolving threats, adversarial settings, and complex interactions.
Manual \ac{ae} struggles to scale with growing submission volumes, complex software-hardware stacks, and deeper methodological flaws.
Advancements in \ac{ai}, particularly concerning \acp{llm}, offer promising solutions to automate and enhance certain tasks in this field.
In this section, we review current \ac{ae} practices and their scalability challenges (\cref{sec:artif-eval-at-sec-conf}), common pitfalls in \ac{ai}-driven cybersecurity research (\cref{sec:comm-mist-cybers}), and emerging \ac{ai}-based automation techniques targeted toward \ac{ai} and cybersecurity (\cref{subsec:background:automation}).

\subsection{Artifact Evaluation at Security Conferences}
\label{sec:artif-eval-at-sec-conf}

Many top-tier cybersecurity conferences have introduced (currently non-mandatory) \ac{ae} into their peer-review process.
\Ac{ae} requires authors to submit the code, datasets, and documentation (typically including a \readme{} with setup and execution instructions) for independent reviewers to verify computational reproducibility.
Consequently, the reviewers can award reproducibility badges if the code is available, runnable, or provides the claimed results \cite{ACM2020Artifact,Olszewskietal2023Get}.

Several papers point out the importance of artifacts and their evaluation in computer science \cite{Timperleyetal2021Understanding,Liuetal2024Research,Hermannetal2020Community,Juristoetal2011The,Gundersenetal2018State,Peng2011Reproducible,Winteretal2022A,Scheitleetal2017Towards,Yildizetal2021ReproducedPapers.org:}, and for cybersecurity in particular \cite{Pennekampetal2021Collaboration,Moultonetal2024Confronting,Uetzetal2021Reproducible}.
This process promotes transparency, encourages best practices in experiment reporting, and accelerates the adoption of the research code in the community and potentially in production environments \cite{Olszewskietal2023Get}.
Despite numerous attempts to formalize the requirements for experiment reporting and implementation description \cite{Kitchenhametal2002Preliminary,Jedlitschkaetal2005Reporting,Timperleyetal2021Understanding}, many researchers emphasize various challenges with reproducing the results \cite{Bajpaietal2017Challenges,Liuetal2024Research,Olszewskietal2023Get,Winteretal2022A,Collbergetal2015Repeatability}.

Despite these benefits, \ac{ae} often demands extensive manual effort and expertise, especially given the growing number of submissions to cybersecurity conferences \cite{Olszewskietal2023Get} and conferences in general \cite{Demetrescuetal2022On}.
Reviewers must resolve complex dependencies and sometimes accommodate specialized hardware requirements \cite{Olszewskietal2023Get}.
Double-blind reviewing intensifies these challenges when anonymization forces the removal of identifying parts of the original code or documentation \cite{Bajpaietal2017Challenges}.

Further analyses reinforce these difficulties: Liu \etal{} \cite{Liuetal2024Research} examine \num{2196} papers and \num{1487} corresponding artifacts submitted between 2017 and 2022 to software engineering venues and find no significant improvement in overall artifact quality, noting in particular that the provided \readme{} files often lack clear instructions and examples.
Olszewski \etal{} \cite{Olszewskietal2023Get} systematically inspect \num{744} \ac{ai}-focused submissions at top security conferences and find that only \num{298} include artifacts.
Out of the available artifacts, only \qty{57}{\percent} provide setup instructions, and not all of these instructions lead to the successful execution \cite{Olszewskietal2023Get}.

Complementing the aforementioned study, we focus on exploring how \acp{llm} can reduce the human workload of \ac{ae} by providing automated support for key steps and comparing our results against this manually established benchmark.

\issue{Conventional \ac{ae} processes no longer scale with rising submission rates and the diversity in utilized software and hardware stacks.}

\subsection{Common Pitfalls in Cybersecurity Research}
\label{sec:comm-mist-cybers}

A rigorous review of a research paper should not only reproduce results but also critically examine the underlying methodology for evaluation and design flaws, complementing \ac{ae}.
Arp \etal{} \cite{Arpetal2022Dos} identify several recurring pitfalls that undermine the scientific validity of cybersecurity submissions.
For example, Sampling Bias or Base Rate Fallacy may lead to overfitting on imbalanced data or inflated detection metrics due to unrealistically high attack rates in the evaluation data \cite{Lones2024Avoiding,Arpetal2022Dos,Pennekampetal2021Collaboration}.
Lab-only evaluations restrict experiments to synthetic environments, failing to capture real-world operational networks' diversity and adaptive strategies \cite{Arpetal2022Dos}.
Conventional \ac{ae}, which focuses primarily on repeating an experiment by rerunning code, often misses these deeper, more foundational issues.
However, they remain relevant from the artifact contribution to the community.

In this work, we thus examine how \acp{llm} can be used to detect textual indicators of these flaws and how to integrate the detection into a (semi-)automated \ac{ae} workflow.

\issue{Detecting methodological flaws in a study is vital to determine its true contribution; however, these flaws are often hard to detect as part of standard reproducibility checks.}

\subsection{\Ac{ai}-Induced Automation Improvements}
\label{subsec:background:automation}

\Acp{llm} have demonstrated strong code understanding, generation, and document analysis capabilities \cite{Zhangetal2025When}.
In cybersecurity, they are already used for vulnerability detection \cite{Guoetal2024Outside}, flagging anomalies or intrusions \cite{Zhangetal2024Large}, and for guiding fuzzing campaigns and penetration tests \cite{Happeetal2023Getting}.
Parallel efforts apply \acp{llm} to peer review:
Numerous authors \cite{Gaoetal2025MMReview:,Caoetal2025CSPaper,Kuznetsovetal2024What,Zhuetal2025DeepReview:,Yeetal2024Are,Idahletal2025OpenReviewer:,Sunetal2024MetaWriter:,Huangetal2025PaperEval:} introduce various techniques to support peer-review processes at academic conferences with \acp{llm}.
While their work provides a foundation for future research on automated academic peer-review systems, they note that some challenges such as susceptibility to adversarial inputs or biases must be resolved before the tools can be widely deployed.
Regarding reproducibility, Bhaskar \cite{Bhaskaretal2024Reproscreener:} introduces an \ac{llm}-based tool to identify reproducibility indicators in \ac{ai}-related papers and their artifacts, achieving better agreement with human judgments when compared to keyword-based approaches.

Despite these advances, a comprehensive automation of \ac{ae}, including execution environment provisioning, subsequent execution, and detection of methodological pitfalls, remains an open challenge.
When complemented with an \Ac{llm}, such a system can substantially reduce \ac{ae} experts' manual workload and improve the consistency and reliability of the \ac{ae} process in cybersecurity research.

Our intuition is that an \ac{llm}-driven toolkit that integrates text-based reproducibility screening, automated setup and execution of artifacts, and the detection of common pitfalls may be marketable given the recent advances in \ac{ai}.
Such a system can substantially reduce \ac{ae} experts' manual workload and improve the consistency and reliability of the \ac{ae} process in cybersecurity research.

\issue{The utility of \ac{ai} for assessing the reproducibility of proposed concepts remains underexplored.}

\section{An \ac{llm}-driven Pipeline to Automate Parts of Artifact Reproducibility Assessments}
\label{sec:repr-pipel}

Having the aforementioned issues in mind and employing recent \ac{ai} developments, we propose an \ac{llm}-driven toolkit that provides automated support for three crucial stages of \ac{ae}:
text-based reproducibility rating (\rate{}, \cf{} \cref{subsec:rating-repr-based}), autonomous execution environment preparation (\prep{}, \cf{} \cref{subsec:using-llms-run}), and methodological-pitfall assessment (\ass{}, \cf{} \cref{subsec:assessing-pitfalls}).
Before introducing the design details of the individual steps, we describe how they can be composed into a modular pipeline (\cref{subsec:design-overview}) to support the manual human peer review.

\subsection{Design Overview}
\label{subsec:design-overview}

\begin{figure}[t]
  \centering
  \includegraphics[width=.85\linewidth]{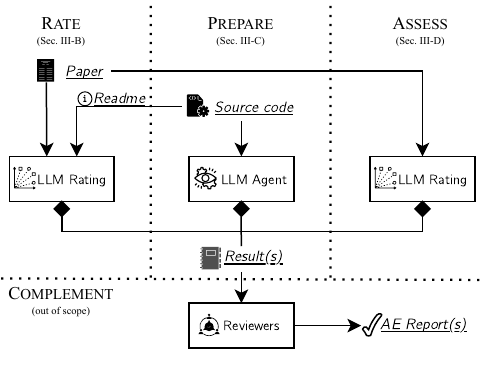}
  \caption{%
    The three pipeline stages require different inputs, and each stage utilizes an \ac{llm}.
    Further, they can be used in combination as desired.
    Any available results can then be fed into the \ac{ae} (\comp{} stage).%
    }
  \label{fig:flow-pipeline}
\end{figure}

\Cref{fig:flow-pipeline} shows the workflow of our pipeline, consisting of three steps that address different parts of \ac{ae} processes.
The steps can be combined as desired for the respective \ac{ae} process, or they can be used independently.
Given this independence, the process can be interrupted at any point, and the generated results can be used or discarded according to the use-case-specific preferences (\eg{}, to exclude submissions with low reproducibility scores from the review).

When using the pipeline in an \ac{ae}, the process could look as follows:
\emph{First}, the \rate{} stage checks how reproducible the contribution appears based on the paper and the \readme{} provided along with the source code.
If the \ac{llm} detects that reproducibility is likely impossible or very challenging, the subsequent stages could, if desired, be canceled.

\emph{Second}, the \prep{} stage attempts to set up the entire research artifact in a fresh container environment to enable its execution using the provided documentation.
The \ac{llm}-based agent used in this stage iteratively issues shell commands to clone the repository, install dependencies, and compile and execute code, while parsing the command's outputs in a feedback loop.
Suppose that the execution fails and the \ac{llm} fails to identify further corrective actions.
In that case, the resulting container and a detailed log of commands and errors are archived for further evaluation by an expert, providing them with first insights.

\emph{Third}, the \ass{} stage focuses on rating the methodological soundness of the submission:
Based on the paper submission, it discovers pitfalls that are common in the design and evaluation of contributions in the field.
The results could contain valuable insights and can improve the feedback on methodology that reviewers are returning eventually.

\emph{Finally}, the generated results of all stages, including any created runtime container (\prep{} stage), can be forwarded to the \ac{ae} reviewers to serve as supplemental material for their ``human'' expert review.
The \comp{} stage is out of scope for this paper, since our goal is to support, streamline, and automate reviewers' work using \ac{ai} rather than to replace their expert judgment.

\subsection{\underline{\rate{}}: Content-Based Reproducibility Ratings}
\label{subsec:rating-repr-based}

\begin{figure}[t]
  \centering
  \includegraphics[width=.6\linewidth]{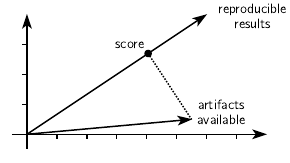}
  \caption{Mapping of concept vectors with a known concept ``reproducible results.''  When measuring a new vector ``artifacts available,'' it is mapped via a projection to the original vector to compute a score.}
  \label{fig:concept-vector-concept}
\end{figure}

Our toolkit's first step, \rate{}, quantifies reproducibility as a semantic direction in an \ac{llm}'s hidden-state space.
We adapt Yang \etal{}'s prior work \cite{Yangetal2024LLM-Measure:} that extracts concept vectors from \acp{llm}' internal states.
By projecting a new text's embedding onto such a concept vector, they quantify how strongly that text represents the respective concept.
The authors demonstrate that this approach yields consistent and valid measures for concepts in social science research contexts.
In our case, we define the concept as \emph{reproducibility} in cybersecurity research.

We begin by crafting two descriptive prompts $p^+$ and $p^-$ that define the opposite poles of our concept: one characterizing that a paper is ``easy to reproduce'' and the other describing that a paper is ``difficult to reproduce.''
These prompts instruct the \ac{llm} to attend to textual cues such as the clarity of methodological descriptions, the presence and quality of installation and execution instructions, as well as the completeness of supplementary materials.

To extract a reproducibility concept vector, we randomly select a set of $n$ probing texts ${t_i, 0 \leq i < n}$ and feed each twice into the \ac{llm}, once under $p^+$ and once under $p^-$.
We extract textual cues from each run in the form of embedding vectors from the final layer $v_i$ of the model, yielding pairs ${(v_i^+, v_i^-)}$.
We then compute ${v_i^\delta := |v_i^+ - v_i^-|}$ for each probe and apply \ac{pca} to the collection ${\{ v_i^\delta : 0 \leq i < n \}}$.
The first principal component serves as our distilled concept vector $\hat{v}$ \cite{Yangetal2024LLM-Measure:}.

To evaluate the reproducibility of a new paper, we obtain its hidden-state embedding $v$ under a neutral prompt and project it onto $\hat{v}$ by computing a dot-product ${s := v \cdot \hat{v} / \|\hat{v}\|}$.
The resulting score $s$ reflects how strongly the paper's text aligns with the distilled reproducibility concept vector constructed from the training dataset.
As the method relies only on hidden-state vectors and \ac{pca}, it is independent of the specific \ac{llm} architecture and can be applied to any model that exposes final-layer embeddings.

\subsection{\underline{\prep{}}: Autonomously Setting up Code}
\label{subsec:using-llms-run}

\begin{figure}[t]
  \centering
  \includegraphics[width=0.7\linewidth]{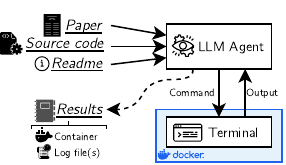}
  \caption{The \ac{llm} agent gets access to the paper, the relevant source code and data, as well as a \readme{} (if available). It then generates commands to execute the code and runs them in a terminal. The outputs are sent back to the agent to determine the next steps.}
  \label{fig:llm-code-execution-flow}
\end{figure}

In the \prep{} stage, we deploy an \ac{llm}-based agent to automate execution environment setup and code execution within a sandboxed environment, as we detail in \cref{fig:llm-code-execution-flow}:
Our agent has full access to a shell and is given
\begin{enumerate*}[label=(\roman*)]
\item the paper,
\item the artifact codebase, and
\item existing documentation, such as a \readme{} file.
\end{enumerate*}
We then prompt it to emit shell commands, which are executed sequentially in a container.

As a first step, we instruct the agent to download any relevant code and datasets required to execute the artifact.
After running a command, we capture the output and send it back to the \ac{llm}, enabling it to diagnose errors such as missing dependencies, version mismatches, or compilation failures, and to generate follow-up commands to resolve errors.
Optionally, the agent may be instructed to output natural-language explanations of each step for human review.

This interactive feedback loop continues until the artifact runs or the agent indicates no further corrective actions are possible.
By isolating each artifact in its own container, we ensure
\begin{enumerate*}[label=(\roman*)]
\item reproducibility by starting from a clean system instance,
\item resource control, such as \ac{gpu} access,
\item clean teardown after the execution, and
\item isolation from other processes running on the host that may otherwise interfere with the execution.
\end{enumerate*}

The final deliverable of this stage is either a runnable container image ready for further analysis by an \ac{ae} expert or a structured error report that pinpoints issues the agent faced.
In the latter case, an expert may manually try to fix the detected problems; nonetheless, the \ac{llm} agent already completed large parts of trial-and-error setups beforehand.

\subsection{\underline{\ass{}}: Identifying Pitfalls in Contributions}
\label{subsec:assessing-pitfalls}

While the previous stages focus on computational reproducibility of the results of a research submission, this stage evaluates the scientific rigor of a submission.
Most importantly, it may enhance the quality of reviews issued by \ac{ae} experts by supporting the detection of otherwise hard-to-notice flaws in the study's methodology and evaluation.
We focus on Arp \etal's \cite{Arpetal2022Dos} taxonomy of ten common pitfalls in \ac{ai}-driven cybersecurity research; however, our approach is conceptually independent of the specifically analyzed pitfalls.
This stage works similarly to the \rate{} stage by independently extracting a concept vector from the underlying \ac{llm} for each of the analyzed pitfalls.

For each of the $m$ analyzed pitfalls, we construct positive and negative prompts that characterize the opposite poles of the respective concept (\ie{}, pitfall present or pitfall not present).
Using the procedure from the \rate{} stage, we derive a unique concept vector for each pitfall individually using a set of training papers.
To assess a new paper, we compute scores $s_i, 0 \leq i < m$ for each pitfall to obtain a feature vector $s := (s_0, \ldots, s_{m-1})$ which we input into a supervised classifier.
The classifier outputs which pitfalls are most likely present.
The report highlights potential design or evaluation flaws, providing reviewers with insights into the submission's potential methodological strengths and weaknesses.

\section{Evaluation}
\label{sec:evaluation}
To demonstrate the effectiveness of our toolkit on real paper submissions, we measure the accuracy and reliability of our individual steps on two expert-annotated datasets:
We employ Olszewski \etal's \cite{Olszewskietal2023Get} dataset of several hundred \ac{ai}-based cybersecurity papers to benchmark \rate{} and \prep{}, and Arp \etal's \cite{Arpetal2022Dos} dataset of \num{30} papers to assess \ass.
We begin by introducing the datasets and our experimental setup in \cref{sec:datasets,sec:setup}, respectively.  We then present results for the combined pipeline and its individual components in \cref{subsec:repr-pipel-1,subsec:eval-in-detail}.

\subsection{Datasets}
\label{sec:datasets}

Reproducibility has no universally accepted quantitative benchmark.
To evaluate our pipeline, we, therefore, rely on two expert-annotated datasets:
first, Olszewski \etal{} \cite{Olszewskietal2023Get} manually assessed the reproducibility of nearly \num{750} \ac{ai}-based security research papers at top-tier conferences.
Second, Arp \etal{} \cite{Arpetal2022Dos} compiled a dataset of 30 papers where they manually record the presence of ten common pitfalls found in studies in cybersecurity.
Next, we introduce them and our experimental setup, including the configured \acp{llm}.

\subsubsection{\olsdata{}}

Olszewski \etal{} \cite{Olszewskietal2023Get} invested over eight person-years to manually check the computational reproducibility of artifacts associated with papers on \ac{ai} in cybersecurity submitted to USENIX Security, ACM CCS, IEEE S\&P, and NDSS between 2013 and 2022.
They assign discrete reproducibility scores to each submission reflecting, \eg{}, the effort required to acquire its code and data, to execute its code, and to reproduce the correct results.
Furthermore, they document the presence of metadata such as links to code repositories, hyperparameter settings, and dataset splitting.

Most notably, the authors find that out of \num{744} analyzed submissions, only \num{298} include artifacts.
Of those artifacts, roughly \qty{57}{\percent} include a \readme{} file that provides instructions for setting up and executing the corresponding code.
The authors only manage to execute \qty{46}{\percent} of the provided artifacts, while only \qty{20}{\percent} of the tested code repositories produce the same results as advertised in the original papers.

For our evaluation of \rate{} and \prep{}, we rely on code repository availability, \readme{} presence, and manual execution success as ground-truth labels.
We only consider the subset of papers where code is available for \prep{} and where, additionally, a \readme{} is available for \rate{}.

\subsubsection{\arpdata{}}
Arp \etal{} \cite{Arpetal2022Dos} manually reviewed \num{30} papers in cybersecurity submitted to top-tier conferences (2011--2021) to identify ten recurring experimental and design pitfalls.
They annotate each paper for the presence or partial presence of each pitfall and whether the authors discuss the flaw in their papers.
Notably, they find that sampling bias affects \qty{90}{\percent} of the analyzed papers, \qty{60}{\percent} rely on an inappropriate threat model, and that other issues, such as base-rate fallacy and lab-only evaluation scenarios, affect a majority of papers.
We rely on the dataset by Arp \etal{} \cite{Arpetal2022Dos} to evaluate the \ass{} step.
While Arp \etal{} also track whether pitfalls are discussed by authors in the text, we only focus on detecting their presence.

\subsection{Experimental Setup}
\label{sec:setup}

We now introduce the hardware, models, and procedures used to implement and evaluate our pipeline and its individual stages.
All \ac{llm}-based components run with fixed prompts and thresholds for binary decisions.

\subsubsection{\rate{} and \ass{}}

For both \rate{} and \ass{}, we run a local instance of \llamathreetwothreebinstruct{}\footnote{\llamathreetwothreebinstructfootnote{}} on a machine equipped with an NVIDIA H-100 Tensor-Core \ac{gpu}.
A prompt template informs the \ac{llm} that it has access to the full paper text and, in the case of \rate{}, a \readme{} file associated with the submission's code artifact.
To derive concept vectors, we fix a random sample of \num{12} papers for \rate{} and \num{10} papers for \ass{} from the respective datasets and run them through the \ac{llm} under the positive and negative prompts.
The remaining papers form the test set.
We compute cutoff scores by optimizing for recall for \rate{} and via logistic regression for \ass{}.

\subsubsection{\prep{}}

Our \ac{llm} agent for \prep{} uses \mbox{OpenAI's} \gptfouromini{}\footnote{\gptfourominifootnote{}} model and interacts with it through the respective web \acs{api}.
Initial experiments with \llamathreetwothreebinstruct reveal that many of the generated commands to run the corresponding artifacts are invalid and that the model quickly runs out of ideas to fix any occurring issues.

For each experiment, the agent spawns a Docker container based on Nvidia's \texttt{cuda} image, which in turn uses Ubuntu 22.04 as its base Linux distribution.
We host the container on a machine equipped with two Intel Xeon Platinum 8160 \acsp{cpu} and two NVIDIA Tesla V-100 \acp{gpu}.
Setting up artifacts that require \acp{gui} or hardware emulations is, unfortunately, not possible in our setup, leading to failed executions of the corresponding code.

\subsection{Reproducibility Pipeline Evaluation}
\label{subsec:repr-pipel-1}

\begin{table}[t]
  \centering
  \caption{Comparison of the output of the reproducibility pipeline with the \olsdata{}. The pipeline correctly classifies almost three-quarters of the examined submissions, providing execution environments for more than \qty{27}{\percent} of all submissions marked as runnable in the ground-truth.}
  \label{tab:pipeline-vs-olszewski}
\begin{tabular}{rrccl}
  \toprule
  \multirow[c]{2}{\widthof{\textsc{Pipeline}}}{\centering{}Total\\126} & & \multicolumn{2}{c}{\olsdata{}} & \\\cmidrule(lr){3-4}
  & & runs & $\lnot$runs & \\\cmidrule(lr){2-5}
  \multirow[c]{2}{\widthof{\textsc{Pipeline}}}{\hfill\textsc{Pipeline}} & runs & $7.14\%$ & $8.73\%$ & $15.87\%$ \\
  & $\lnot$runs & $19.05\%$ & $65.08\%$ & $84.13\%$ \\\cmidrule(lr){2-5}
  & & $26.19\%$ & $73.81\%$ & \\
  \bottomrule
\end{tabular}\vspace{1ex}
\begin{tabular}{rlrlrl}
  \textbf{Accuracy:} & $\mathbf{72.22}\textbf{\%}$ & Precision: & $45.00\%$ & Recall: & $27.27\%$
\end{tabular}

\end{table}

\begin{table}[t]
  \centering
  \caption{Comparison of the output of the \rate{} stage with the \olsdata{}. The approach correctly classifies almost all submissions marked as runnable in the ground-truth.}
  \label{tab:rate-vs-olszewski}
\begin{tabular}{rrccl}
  \toprule
  \multirow[c]{2}{\widthof{\rate{}}}{\centering{}Total\\130} & & \multicolumn{2}{c}{\olsdata{}} & \\\cmidrule(lr){3-4}
  & & runs & $\lnot$runs & \\\cmidrule(lr){2-5}
  \multirow[c]{2}{\widthof{\rate{}}}{\hfill\rate{}} & runs & $40.77\%$ & $54.62\%$ & $95.38\%$ \\
  & $\lnot$runs & $2.31\%$ & $2.31\%$ & $4.62\%$ \\\cmidrule(lr){2-5}
  & & $43.08\%$ & $56.92\%$ & \\
  \bottomrule
\end{tabular}\vspace{1ex}
\begin{tabular}{rlrlrl}
  Accuracy: & $43.08\%$ & Precision: & $42.74\%$ & \textbf{Recall:} & $\mathbf{94.64}\textbf{\%}$
\end{tabular}

\end{table}

\begin{table}[t]
  \centering
  \caption{Comparison of the output of the \prep{} stage with the \olsdata{}. The agent automatically sets up ready-to-use execution environments for almost \qty{29}{\percent} of all submissions marked as runnable in the ground-truth.}
  \label{tab:prep-vs-olszewski}
\begin{tabular}{rrccl}
  \toprule
  \multirow[c]{2}{\widthof{\prep{}}}{\centering{}Total\\311} & & \multicolumn{2}{c}{\olsdata{}} & \\\cmidrule(lr){3-4}
  & & runs & $\lnot$runs & \\\cmidrule(lr){2-5}
  \multirow[c]{2}{\widthof{\prep{}}}{\hfill\prep{}} & runs & $7.40\%$ & $14.79\%$ & $22.19\%$ \\
  & $\lnot$runs & $18.97\%$ & $58.84\%$ & $77.81\%$ \\\cmidrule(lr){2-5}
  & & $26.37\%$ & $73.63\%$ & \\
  \bottomrule
\end{tabular}\vspace{1ex}
\begin{tabular}{rlrlrl}
  \textbf{Accuracy:} & $\mathbf{66.24}\textbf{\%}$ & Precision: & $33.33\%$ & Recall: & $28.05\%$
\end{tabular}

\end{table}

\Cref{tab:pipeline-vs-olszewski} illustrates the overall performance of our pipeline, \ie{}, the combination of the \rate{} and \prep{} stages.
We consider the intersection of papers from Olszewski \etal's \cite{Olszewskietal2023Get} dataset in both stages individually.
Overall, our system correctly assesses whether an artifact's code can be executed without major effort in more than \qty{72}{\percent} of cases.

Although only about \qty{7}{\percent} of \emph{all} attempted artifacts are fully containerized and executed by the pipeline, this performance corresponds to provisioning runnable environments for roughly \qty{28}{\percent} of the papers that Olszewski \etal{} \cite{Olszewskietal2023Get} manage to execute out of the box, \ie{}, using only the instructions in the corresponding \readme{} files.
Only about \qty{7}{\percent} of papers are misclassified as non-runnable when they, in fact, can run out of the box according to Olszewski \etal{} \cite{Olszewskietal2023Get}.
These false negatives are often induced by our Docker environment, which cannot emulate special hardware, or by the agent, which cannot fix them without external inputs, \eg{}, if the link to the sources from the dataset leads to an informative website instead of a Git repository.
In any case, the agent provides a reason for failure that human \ac{ae} experts can use to try to fix the remaining issues manually.

The true negative rate of the pipeline exceeds \qty{85}{\percent}, meaning that our pipeline reliably filters out non-runnable submissions.
These numbers underline that the proposed pipeline can indeed save reviewers from spending valuable time otherwise spent on setting up artifacts using trial-and-error, which is a process that is comparably easy to automate.
By prepending the \ass{} stage, submissions deemed unlikely to be reproducible can even be discarded before entering the more costly \prep{} stage, saving valuable computational resources.
The results of the pipeline are shared with an \ac{ae} expert in the \comp{} stage, who can then decide on, \eg{}, whether to award a reproducibility badge to the given submission.

\subsection{Detailed Evaluation Results}
\label{subsec:eval-in-detail}

In this section, we give a more detailed overview of the classification results of the individual stages of our pipeline.
We highlight how \rate{} reliably forwards nearly all runnable artifacts to the next stage, how \prep{} autonomously provides numerous execution environments, and how \ass{} detects methodological flaws with high accuracy.

\subsubsection{\rate{}}
\label{subsubsec:rating-repr-based-1}

This stage aims to find papers whose code is likely not runnable and discard them early, before wasting computational resources and time setting up the code.
For the evaluation of this stage, we consider only papers where the code as well as a \readme{} file are available.

\Cref{tab:rate-vs-olszewski} compares this stage's classification results to the \olsdata{} \cite{Olszewskietal2023Get} dataset.
The high recall of almost \qty{95}{\percent} indicates that almost all papers with runnable code are selected to move to the next stage (the false negative rate is just over \qty{6}{\percent}).
In fact, fewer than \qty{3}{\percent} of all analyzed papers are misclassified as not runnable.

This result makes the step ideal as a first stage for our pipeline: if a paper is deemed not runnable, no computational resources need to be spent to try and execute the respective code.
Instead, \ac{ae} experts are given the result of the stage.
If they feel the results can be reproduced after all, the experts can still manually feed the respective code and paper into the \prep{} stage.
Given the small number of false negatives, only a few papers are discarded early in the pipeline and not automatically examined for execution.

\subsubsection{\prep{}}
\label{subsubsec:agent-based-repr}

Unlike the previous stage, this stage's goal is to automatically set up execution environments to enable experts to quickly run a paper's code and manually assess the validity of the reproduced results.
We consider 311 papers from the dataset since our agent does not explicitly require the presence of a \readme{} file; instead, it can autonomously analyze the code repository structure and, \eg{}, try to compile and execute relevant files.
All papers that have been evaluated in the \rate{} stage are also analyzed in this stage.

In \cref{tab:prep-vs-olszewski}, we summarize the results of the classification, which show that the agent yields a moderately high accuracy of more than \qty{66}{\percent}.
Notably, this stage alone reliably eliminates the need for experts to manually set up execution environments for papers whose code is not runnable for almost \qty{60}{\percent} of the analyzed papers.
The false negatives are often induced by limitations of our execution environment (\cf{} \cref{sec:setup}):
Even though our \ac{llm} agent can effectively execute terminal commands, some artifacts require access to graphical desktop environments to, \eg{}, run Internet browsers, which is out of the scope of our experiments.

\subsubsection{\ass{}}
\label{subsubsec:pitfalls}

Given the relatively small size of the dataset from the \arpdata{} \cite{Arpetal2022Dos}, \ie{}, \num{30} papers, we cannot analyze the pitfalls on \emph{sampling bias} (P1) and \emph{data snooping} (P3).
This limitation is due to our training process requiring at least \num{5} papers per category ``pitfall present'' and ``pitfall not present.''
For the remaining pitfalls, our evaluation yields promising results.
Except for the pitfall on \emph{biased parameters} (P5), the classifier has an accuracy between \qty{90}{\percent} and \qty{100}{\percent}.
$F_1$ scores are between \num{.92} and \num{1}, and $F_2$ scores between \num{.97} and \num{1}, indicating an accurate response given by our approach.
For (P5), our approach performs almost like a random predictor.
However, Arp \etal{} \cite{Arpetal2022Dos} classify most papers as ``unclear from text'' for this category.
We presume that a larger and more representative dataset would fix this problem.
The remaining seven pitfalls can be accurately detected using only a small human-annotated dataset.
Overall, we conclude that \ass{} is well-suited for detecting common known pitfalls in security-related research papers on \ac{ai}.

\section{Discussion and Future Work}
\label{sec:discussion}

Our results show that an \ac{llm}-driven toolkit can reliably filter out non-runnable submissions, autonomously provide execution environments for submitted artifacts, and accurately flag common methodological pitfalls in cybersecurity research.
In the following, we discuss our findings in more detail while also addressing limitations of our design, implementation, and evaluation, and proposing directions for future research on the topic.
Further, we complement this discussion of findings with a brief overview of lessons learned during our research activities in \Cref{app:sec:lessons-learned}.

\subsection{Individual Findings}
\label{subsec:individual-findings}

Given the overall results of the toolkit, we reflect upon the individual components' strengths and limitations.
Furthermore, we outline targeted directions for future enhancements.

\subsubsection{\rate{}}

This stage already yields promising results despite the \ac{llm} used for this purpose not being fine-tuned to the given task.
Instead, the training data is given to the \ac{llm} as a prompt.
Future work may evaluate whether fine-tuning an \ac{llm} improves the quantification of the concept of artifact reproducibility within the model to generate more precise and consistent concept vectors.
However, this change would require a large amount of training data, which is unavailable to us at the time of writing.
This training data could, for example, be collected as part of a shadow \ac{ae} conducted to evaluate the pipeline further, as suggested in \cref{subsec:general-findings}.

\subsubsection{\prep{}}

While this stage automatically creates sandboxed execution environments for many paper artifacts, with the currently used execution environment, we are still unable to handle all submissions correctly.
This situation is partly due to technical limitations, \eg{}, the lack of a desktop environment or specialized hardware required for some evaluations.
The former could be solved by adding \ac{gui} interaction support to the agent, \eg{}, using UI-TARS \cite{Qinetal2025UI-TARS:}.
The latter can be solved by providing a more diverse hardware setup for the stage; this improvement, however, exceeds the scope of this paper, as our goal is to show the general feasibility of the approach.

\subsubsection{\ass{}}

Our evaluation of this stage shows that the detection of pitfalls in cybersecurity papers on \ac{ai} performs very reliably.
However, the small size of the evaluation dataset poses limitations to our evaluation.
We propose to re-evaluate this stage on a more exhaustive dataset.
The creation of such a dataset is, however, infeasible within the scope of this paper.

\subsection{General Findings and Future Directions}
\label{subsec:general-findings}

Our evaluation shows that our tools, when combined into a pipeline, can provide significant support in the \ac{ae} process conducted at security conferences.
It provides a first step into automating this process by detecting submissions without reproducible artifacts and autonomously preparing their execution to enable \ac{ae} experts to more quickly assess the validity of the reproduced results.
Hence, we provide means to significantly boost the scalability of the process, particularly as we facilitate the tedious task of setting up code environments for performing the evaluations.

\subsubsection{Open Questions}
\label{sec:open-questions}

Despite the potential highlighted in our evaluation, we identify several open questions:
\begin{enumerate*}[label=(\roman*)]
\item Better understanding in detail how ``perfect'' prompts could look like for the different approaches in our toolkit.
\item Further comparing different underlying \acp{llm}, as different models may be better in finding and understanding certain concepts or performing certain tasks, in particular, depending on the model size.
\item Assessing the security risks of applying our pipeline in practice, \eg{}, regarding the execution of arbitrary code in the artifacts, as well as better understanding the implications for intellectual property fed to closed-source commercial models.
\end{enumerate*}
Concerning the last question, \prep{} already provides execution environments that are sandboxed in individual Docker containers.
However, access to hardware components such as \acp{gpu} or other specialized devices may impose additional risks on the system.

\subsubsection{Integration into Peer-Review}
\label{sec:integr-into-peer}

After improving the techniques for automatically assessing artifact reproducibility proposed in this paper, future work may integrate them directly into the review process at cybersecurity conferences.
Given more general training data, the process could also be integrated into conferences in other fields.
Currently, artifact reproducibility checks are often only performed for accepted papers, \ie{}, after the review process is completed \cite{Athanassoulisetal2022Artifacts}.
Automated reproducibility checks would allow checking a large number of submissions even before issuing an acceptance.
While some authors might be concerned with participating in a potentially biased or low-quality \acp{ae}, an \ac{ai}-assisted pipeline may increase their trust in the process and, in turn, improve their willingness to participate.
We believe that assessing the usability of the proposed pipeline workflow in the form of shadow \ac{ae} is a good next step for assessing its maturity.

Integrating our tool into peer review requires addressing manipulation risks, \eg{}, prompt injection. The attack surface is limited in \rate{} and \ass{}, as outputs derive from concept vectors extracted from the \ac{llm}'s hidden states rather than direct generation from paper text.
Reviewers read the paper before or in parallel to the partially automated \ac{ae}, aiding detection of any injections.
In \prep{}, injections may occur in \readme{} files or code comments; however, sandboxed execution and expert assessment of results render their impact negligible.
We expect such malicious acts to be rare due to scientific integrity norms and penalized when discovered.

\subsubsection{Future Evaluations}
\label{sec:future-evaluations}

Finally, future work may expand our research by applying the techniques to papers from different fields.
We limit our evaluation to papers in these domains due to the availability of an exhaustive dataset.
We believe the approach easily generalizes to other topic areas as it does not directly depend on the contents of the evaluated works.
Most importantly, we show that \ac{ai} is a promising tool to employ in \ac{ae} with a great potential to complement the process to improve its quality, scalability, and thus sustainability.

\subsection{Lessons Learned}
\label{app:sec:lessons-learned}

During the development of our pipeline, we noticed several unexpected behaviors across interactive handling, execution, verification, and contextual reasoning.
For example, in the \prep{} stage, the agent reports that it cannot engage with an interactive editor such as \texttt{nano} for one experiment.
While it could have proposed an alternative non-interactive solution, \eg{}, using \texttt{sed}, the \ac{llm} did not have this idea, revealing a gap in its problem-solving repertoire.
We suggest evaluating this behavior with more powerful models to assess whether this problem can be resolved.

Furthermore, in one experimental run during our implementation, the agent proposed to comment out an entire program to make it run successfully---returning in an inaccurate assessment.
However, this change results in the program not performing any computations or providing any outputs.
We worked around this issue by adapting the prompts given to the \ac{llm}, highlighting the importance of carefully designed prompts and validation.

In the \rate{} stage, we notice that the \ac{llm}'s grasp of the concept of reproducibility is less accurate than that of the different pitfalls analyzed in the \ass{} stage.
This result may be induced by the training data of the utilized model:
Reproducibility is a niche topic, and today's models are likely not trained on much input that covers this concept.

Simultaneously, even on powerful hardware, local models execute much more slowly than commercial models like \gptfouromini, which are heavily optimized for mass use.
In particular, to protect the privacy and confidentiality of submissions during the (confidential) \ac{ae} process, we propose that future work focuses on optimizing \acp{llm} specifically for the use case of artifact reproducibility assessment.

Overall, we have learned that \acp{llm} constitute a powerful tool that has the potential to substantially complement and improve the \ac{ae} process at scientific conferences.
They can be employed to automate tedious and repetitive tasks while simultaneously streamlining the whole process to help provide more consistent and high-quality feedback to authors.

\section{Conclusion}
\label{sec:conclusion}

Ensuring the reproducibility of research artifacts in cybersecurity is crucial in science to validate the potential for further use of given experimental and methodological results.
It narrows the gap between experiments and simulations, as well as real-world deployments, since stakeholders can better assess the suitability of the approaches for their systems.
Currently, some scientific conferences perform time-consuming manual artifact evaluations to assess whether the contributions of the submitted works are reproducible.
We propose an \ac{llm}-based toolkit that enhances the automation potential of otherwise manual and time-consuming artifact assessments.
Our evaluation shows that, when combining the tools into a pipeline, a majority of submissions without runnable artifacts are automatically discarded.
At the same time, execution environments are generated for many submissions with runnable code.
We propose to integrate such a pipeline into the \acf{ae} process of conferences to incentivize researchers to deliver reproducible results.
Furthermore, this change has potential to unburden reviewers by automating a time-consuming part of the review work, with the objective of improving the sustainability of the \ac{ae} process.

\section*{Acknowledgments}

This work was funded by the \acf{bmftr} in Germany under the grant number 16KIS2251 of the SUSTAINET-guardian project.
The responsibility for the content of this publication lies with the authors.
The authors thank Daniel Arp for supporting the pitfall evaluation (\ass{} stage), which builds upon the survey data by Arp \etal{} \cite{Arpetal2022Dos}.

\bibliographystyle{template/IEEEtran}
\bibliography{paper}

\end{document}